\documentclass{ws-ijgmmp}

\begin{document}


%
\catchline{}{}{}{}{}
%

\title{Matrix 3-Lie superalgebras and BRST supersymmetry}

\author{VIKTOR ABRAMOV}

\address{Institute of Mathematics and Statistics, University of Tartu, J. Liivi 2 - 602\\
Tartu 50409, Estonia\\
\email{viktor.abramov@ut.ee}}



\maketitle

\begin{history}
\received{(April 2017)}
\revised{(Day Month 2017)}
\end{history}

\begin{abstract}
Given a matrix Lie algebra one can construct the 3-Lie algebra by means of the trace of a matrix. In the present paper we show that this approach can be extended to the infinite-dimensional Lie algebra of vector fields on a manifold if instead of the trace of a matrix we consider a differential 1-form which satisfies certain conditions. Then we show that the same approach can be extended to matrix Lie superalgebras $\frak{gl}\,(m,n)$ if instead of the trace of a matrix we make use of the super trace of a matrix. It is proved that a graded triple commutator of matrices constructed with the help of the graded commutator and the super trace satisfies a graded ternary Filippov-Jacobi identity. In two particular cases of $\frak{gl}\,(1,2)$ and $\frak{gl}(2,2)$ we show that the Pauli and Dirac matrices generate the matrix 3-Lie superalgebras, and we find the non-trivial graded triple commutators of these algebras. We propose a Clifford algebra approach to 3-Lie superalgebras induced by Lie superalgebras. We also discuss an application of matrix 3-Lie superalgebras in BRST-formalism.
\end{abstract}


\section{Introduction}
\noindent
A concept of $n$-Lie algebra, where $n\geq 2$, was introduced and studied by V.T. Filippov in \cite{Filippov}. This structure turned out to be fruitful, and one of reasons for interest in this structure is its connection with a generalization of classical mechanics, where a ternary Nambu bracket replaces the classical binary Poisson bracket, proposed by Y. Nambu in \cite{Nambu}. The question of quantization of the Nambu bracket turned out to be difficult, and the quantization problem of the Nambu bracket is studied in \cite{Awata-Li-Minic-Yaneya}, \cite{Dito-Flato-Sternheimer}, \cite{Takhtajan}. Recent developments in this field of research show that 3-Lie algebras play an important role in M-theory generalization of the Nahm equation \cite{Lambert},\cite{Palmer}.

\vskip.3cm
\noindent
In the paper \cite{Awata-Li-Minic-Yaneya} devoted to the problem of quantization of a triple Nambu bracket the authors show that given a matrix Lie algebra $\frak{gl}\,(n)$ of matrices on $n$th order one can define the triple commutator by means of the usual binary commutator of two matrices and the trace of a matrix as follows:
\begin{equation}
[A,B,C]=\mbox{Tr}\,A\;[B,C]+\mbox{Tr}\,B\;[C,A]+\mbox{Tr}\,C\;[A,B],
\label{introduction formula 1}
\end{equation}
where $A,B,C\in {\frak{gl}}\,(n)$. Then the authors prove that this triple commutator satisfy the ternary Filippov-Jacobi identity and hence a matrix Lie algebra $\frak{gl}\,(n)$ induces the 3-Lie algebra. The cohomologies of induced 3-Lie algebras are studied in \cite{Makhlouf}, \cite{silvestrov2014}. Our aim in the present paper is to show that the method of induced 3-Lie algebras can be extended to infinite-dimensional Lie algebra of vector fields of a smooth manifold and to matrix 3-Lie superalgebras. In the case of the Lie algebra of vector fields of a manifold we define a triple commutator of three vector fields by a formula analogous to (\ref{introduction formula 1}), where the trace of a matrix is replaced by a differential 1-form $\omega$. Then it can be proved that a triple commutator of vector fields satisfies the Filippov-Jacobi identity if a differential 1-form satisfies two conditions
\begin{eqnarray}
\omega(X)\,Y(\omega(Z)) = \omega(Y)\,X(\omega(Z)),\quad\;\;\;\omega\wedge d\omega = 0,
\end{eqnarray}
where $X,Y,Z$ are vector fields.

\vskip.3cm
\noindent
In the case of a matrix Lie superalgebra $\frak{glt}(m,n)$ we define a graded triple commutator of three matrices by a formula similar to (\ref{introduction formula 1}), where instead of the trace of a matrix we use the super trace and graded binary commutator \cite{abramov2014}, \cite{Abramov-Lätt}. We prove that a graded triple commutator of matrices satisfies the graded Filippov-Jacobi identity. Hence a matrix Lie superalgebra $\frak{gl}\,(m,n)$ induces the matrix 3-Lie superalgebra, which we denote by $\frak{glt}(m,n)$. Then we consider two particular 3-Lie superalgebras $\frak{glt}\,(1,1), \frak{glt}(2,2)$. In the case of the first one we take the Pauli matrices as the generators and find all non-trivial graded triple commutators of the Pauli matrices. Then we take the Dirac matrices as the generators of the second 3-Lie superalgebra and also find all non-trivial graded triple commutators of the Dirac matrices. Finally we consider a general case of 3-Lie superalgebra $\frak{glt}(n,n)$ and apply a supermodule of spinors over a Clifford algebra \cite{Quillen-Mathai} in order to find all non-trivial graded triple commutators of $\frak{glt}(n,n)$.

\vskip.3cm
\noindent
We conclude the present paper by discussion of possible application of a matrix 3-Lie superalgebra $\frak{glt}(m,n)$ in a gauge field theory. We propose an analog of supersymmetry transformation $\delta$ constructed by means of a graded triple commutator. We show that the structure of this supersymmetry transformation is similar to usual BRST-supersymmetry operator \cite{Faddeev-Slavnov} with additional new term which appears due to a ternary structure of graded commutator. Our analog of supersymmetry transformation depends on two "fields", where one is a "bosonic field" and the second is a "fermionic field", which play a role of parameters of supersymmetry transformation. We show that a fermionic field can be identified with a ghost field and a bosonic field with an auxiliary field $B$ of BRST-supersymmetry operator. We find a condition for $\delta^2=0$.
\section{3-Lie algebra of vector fields}
We remind that a vector space $\frak g$ is said to be a Lie algebra if $\frak g$ is endowed with a binary Lie bracket $[\;,\;]:\frak g\times\frak g\to \frak g$, which is skew-symmetric $[a,b]=-[b,a]$ and satisfies the Jacobi identity
$$
[a,[b,c]]+[b,[c,a]]+[c,[a,b]]=0.
$$
The Jacoby identity can be written in the form
\begin{equation}
[a,[b,c]]=[[a,b],c]+[b,[a,c]],
\label{Jacoby identity derivation binary}
\end{equation}
which shows that Lie bracket with an element $a$ applied to Lie bracket $[b,c]$ can be viewed as a derivation.

\vskip.3cm
\noindent
One way to generalize a concept of Lie algebra is to increase the number of arguments in Lie bracket, i.e. to consider $n$-ary Lie bracket. In this case the binary Jacoby identity (\ref{Jacoby identity derivation binary}) suggests that an analog of Jacoby identity in the case of $n$-ary Lie bracket can be obtained if one requires for $n$-ary Lie bracket to be a derivation. This way of generalization was proposed by V.T. Filippov in \cite{Filippov} and later this generalization of a concept of Lie algebra turned out to be very useful in generalized Nambu mechanics \cite{Takhtajan}. An $n$-Lie algebra $\frak g$ is a vector space equipped with a $n$-ary Lie bracket $[\;.\;,\;\ldots\;,\;.\;]:\frak g\times\;\ldots\;\times\frak g\;\mbox{($n$ {times})}\to \frak g$,
which is skew-symmetric, i.e. an $n$-ary Lie bracket changes the sign if one interchanges the positions of any two elements insight $n$-ary Lie bracket
$$
[a_1,\ldots,a_i,\ldots,a_j,\ldots,a_n]=-[a_1,\ldots,a_j,\ldots,a_i,\ldots,a_n],
$$
and satisfies the $n$-ary Filippov-Jacobi identity
\begin{eqnarray}
[a_1,\ldots,a_{n-1},[b_1,\ldots,b_n]] =\sum_{k=1}^n [b_1,\ldots,[a_1,\ldots,a_{n-1},b_k],\ldots,b_n].
\label{n-ary Filippov identity}
\end{eqnarray}
Particularly a 3-Lie algebra is a vector space $\frak g$ equipped with a ternary Lie bracket $[\;,\;,\;]:\frak g\times\frak g\times\frak g\to \frak g$, which satisfies
\begin{eqnarray}
[a,b,c]=[b,c,a]=[c,a,b],\;\;\;[a,b,c]=-[b,a,c]=-[a,c,b]=-[c,b,a],
\label{symmetries of ternary commutator 2}
\end{eqnarray}
and the ternary Filippov-Jacobi identity
\begin{eqnarray}
 [a,b[c,d,e]] = [[a,b,c],d,e]+[c,[a,b,d],e]+[c,d,[a,b,e]].
\label{ternary Filippov identity}
\end{eqnarray}
Let $e_1,e_2,\ldots,e_n$ be a basis for a 3-Lie algebra $\frak g$. The structural constants of 3-Lie algebra $\frak g$ are defined by
\begin{equation}
[e_i,e_j,e_k]=t_{ijk}^l\,e_l.
\end{equation}
From the symmetries of ternary commutator (\ref{symmetries of ternary commutator 2}) and the ternary Filippov-Jacobi identity (\ref{ternary Filippov identity}) it follows that the structural constants of 3-Lie algebra $\frak g$ satisfy
$$
t^l_{ijk}=t^l_{jki}=t^l_{kij},\;\;t^l_{ijk}=-t^l_{jik}=-t^l_{ikj}=-t^l_{kji},
$$
and
\begin{equation}
t^m_{ijk}\,t^k_{rps}+t^m_{psk}\,t^k_{rji}+t^m_{rsk}\,t^k_{jpi}+t^m_{rpk}\,t^k_{jis}=0.
\end{equation}

\vskip.3cm
\noindent
In \cite{Awata-Li-Minic-Yaneya} the authors showed that given a matrix Lie algebra $\frak{gl}(n)$ of square matrices of $n$th order one can construct a 3-Lie algebra by means of trace of a matrix. We briefly remind the method proposed in \cite{Awata-Li-Minic-Yaneya}. If $A,B,C\in \frak{gl}(n)$ are three square matrices of $n$th order then one can define their ternary commutator by
\begin{eqnarray}
[A,B,C] =\mbox{Tr}(A)\;[B,C]+\mbox{Tr}(B)\;[C,A]+\mbox{Tr}(C)\;[A,B].
\label{ternary commutator with trace}
\end{eqnarray}
It is proved in \cite{Awata-Li-Minic-Yaneya} that ternary commutator (\ref{ternary commutator with trace}) satisfies the ternary Filippov-Jacobi identity, and this makes a matrix Lie algebra $\frak{gl}(n)$ the 3-Lie algebra. Later in \cite{silvestrov2014} this method was extended to $m$-Lie algebras with an analog of a trace and the cohomologies of induced $(m+1)$-Lie algebras were studied. In this paper we extend the method of induced $3$-Lie algebras of \cite{Awata-Li-Minic-Yaneya} to Lie algebra of vector fields of a smooth finite dimensional manifold obtaining an infinite-dimensional 3-Lie algebra of vector fields.

\vskip.3cm
\noindent
Let $M$ be a smooth $n$-dimensional manifold, $\frak{D}(M)$ be the Lie algebra of vector fields of $M$, and $\Omega^k(M)$ be the space of differential forms of $k$th degree. Let $\omega\in \Omega^1(M)$ be a differential 1-form and $d:\Omega^k(M)\to\Omega^{k+1}(M)$ be the exterior differential. We define the ternary commutator of three vector fields $X,Y,Z\in \frak{D}(M)$ by
\begin{equation}
[X,Y,Z]=\omega(X)\;[Y,Z]+\omega(Y)\;[Z,X]+\omega(Z)\;[X,Y].
\label{ternary commutator with form}
\end{equation}
\begin{theorem}
The ternary commutator (\ref{ternary commutator with form}) satisfies the ternary Filippov-Jacobi identity (\ref{ternary Filippov identity}) if a 1-form $\omega$ satisfies two conditions:
\begin{enumerate}
\item
$\omega(X)\,Y(\omega(Z))=\omega(Y)\,X(\omega(Z)),\;\;X,Y,Z\in \frak{D}(M)$,
\item
$\omega\wedge d\omega=0$.
\end{enumerate}
\label{Theorem 1}
\end{theorem}
\noindent
We can find an equivalent form of the second condition by making use of the formula
\begin{equation}
d\omega(X,Y)=X(\omega(Y))-Y(\omega(X))-\omega([X,Y]).\label{exterior differential}
\end{equation}
Indeed we have
\begin{eqnarray}
\omega\wedge d\omega(X,Y,Z) = \omega(X)\,d\omega(Y,Z)-\omega(Y)\,d\omega(X,Z)
                             +\omega(Z)\,d\omega(Y,Z).
\label{omega d omega}
\end{eqnarray}
Now substituting the formula for exterior differential (\ref{exterior differential}) into $\omega$ in (\ref{omega d omega}) and making use of the first condition of Theorem \ref{Theorem 1} we obtain the equivalent form of the second condition
\begin{equation}
\omega(X)\,\omega([Y,Z])+\omega(Y)\,\omega([Z,X])+\omega(Z)\,\omega([X,Y])=0.
\end{equation}

\vskip.3cm
\noindent
We can apply Theorem \ref{Theorem 1} to finite dimensional Lie groups. Let $G$ be a finite dimensional Lie group $\mbox{dim}\,G=r$,
$\frak g$ be its Lie algebra of left-invariant vector fields, $X_1,X_2,\ldots,X_n$ be a basis for the Lie algebra $\frak g$, and $\theta^1,\theta^2,\ldots,\theta^n$ be the dual basis for the algebra of left-invariant forms. Let $\omega$ be a left-invariant form of $G$. Then for any two left-invariant vector fields $X,Y$ we have $X(\omega(Y))=0$. Hence any left-invariant 1-form $\omega$ satisfies the first condition of Theorem \ref{Theorem 1}. Thus a left-invariant 1-form $\omega$ can be used to construct the 3-Lie algebra structure on a Lie algebra $\frak g$ if it satisfies $\omega\wedge d\omega=0$. Particularly any closed left-invariant 1-form $d\omega=0$ can be used in (\ref{ternary commutator with form}) to construct a ternary commutator, which satisfies the ternary Filippov-Jacobi identity.

\vskip.3cm
\noindent
As an example let us consider a matrix Lie group $\mbox{GL}\,(n)$ of invertible square matrices of $n$th order and its Lie algebra $\frak{gl}\,(n)$ of all square matrices of $n$th order. We can choose the basis $E^k_l$, where $(E^k_l)^i_j=\delta^k_j\,\delta^i_l$, for the Lie algebra $\frak{gl}\,(n)$. Then
$$
[E^i_j,E^k_l]=\delta^i_l\,E^k_j-\delta^k_j\,E^i_l.
$$
Let $\theta=(\theta^i_j)$ be the matrix of dual basis for $\frak{gl}^*\,(n)$, where $\theta^i_j(E^k_l)=\delta^i_l\,\delta_j^k$. Then
$$
d\theta^i_j=\theta^i_l\wedge\theta^l_j,
$$
or in matrix form $d\theta=\theta\wedge\theta$.
The 1-form $\omega=\mbox{Tr}\,\theta$ is closed $d\omega=0$, because
$$
d\omega=d(\mbox{Tr}\,\theta)=\mbox{Tr}\,d\theta=\mbox{Tr}\,\theta\wedge\theta,
$$
and $\mbox{Tr}\,\theta\wedge\theta=0$. If $A=(A^i_j)\in \frak{gl}\,(n)$ is any matrix of $n$th order then it is easy to see that $\omega(A)=\mbox{Tr}\, A$ and we proved as the special case of Theorem \ref{Theorem 1} that the ternary commutator (\ref{ternary commutator with trace}) satisfies the ternary Filippov-Jacobi identity.
\section{Matrix 3-Lie superalgebras $\frak{glt}(m,n)$}
A concept of Lie algebra can be extended to $\mathbb Z_2$-graded structures, and this extension is called a Lie superalgebra. Similarly the notion of $n$-Lie algebra can be extended to $\mathbb Z_2$-graded case, and we call this extension a $n$-Lie superalgebra. In this section we remind a definition of 3-Lie superalgebra and show that the method of constructing matrix 3-Lie algebras by means of the binary commutator of two matrices and the trace can be applied to construct a matrix 3-Lie superalgebra if we use the graded commutator of two matrices and the supertrace. We prove that a graded ternary commutator of three matrices analogous to ternary commutator (\ref{ternary commutator with trace}) of previous section, where we use graded commutator and supertrace instead of commutator and trace, satisfies graded ternary Filippov-Jacobi identity.

\vskip.3cm
\noindent
A concept of Lie algebra can be extended to $\mathbb Z_2$-graded case by means of a notion of Lie superalgebra. Let $\frak g=\frak g_0\oplus\frak g_1$ be a $\mathbb Z_2$-graded vector space or super vector space. The degree of a homogeneous element $a$ of $\frak g$ will be denoted by $|a|$, and, as usual, the elements of subspace $\frak g_0$ will be called even elements and the elements of $\frak g_1$ will be called odd elements of super vector space $\frak g$. A super vector space $\frak g$ is a Lie superalgebra if it is equipped with a graded Lie bracket $[\;,\;]:\frak g\times\frak g\to \frak g$ which is graded skew-symmetric and satisfies the graded Jacobi identity
\begin{eqnarray}
(-1)^{{|a|}{|c|}}[a,[b,c]]+(-1)^{{|a|}{|b|}}[b,[c,a]]+(-1)^{{|b|}{|c|}}[c,[a,b]]=0.\nonumber
\end{eqnarray}
The graded Jacobi identity can be written in the form
\begin{equation}
[c,[a,b]]=[[c,a],b]+(-1)^{|a||c|}[a,[c,b]].
\label{graded binary Jacobi identity}
\end{equation}
We remind that a graded Lie bracket means that $[\frak g_\alpha,\frak g_\beta]\subseteq\frak g_{\alpha+\beta},\;\alpha,\beta\in {\mathbb Z}_2$ or, equivalently, the degree of $[a,b]$, where $a,b$ are homogeneous elements of $\frak g$, is equal to the sum of degrees of $a$ and $b$, i.e. $|[a,b]|=|a|+|b|$. In what follows we will denote the sum of degrees of two homogeneous elements $a,b$ by $|ab|$ that is $|a|+|b|=|ab|$. A graded skew-symmetry of graded Lie bracket means that for any two homogeneous elements $a,b\in \frak g$ we have
\begin{equation}
[a,b]=-(-1)^{{|a|}{|b|}}[b,a].
\end{equation}

\vskip.3cm
\noindent
A super vector space $\frak g =\frak g_0\oplus\frak g_1$ is said to be a 3-Lie superalgebra if it is endowed with a graded triple Lie bracket $[\;.\;,\;.\;,\;.\;]:\frak g\times\frak g\times \frak g\to \frak g$, i.e.
$$
[\mathfrak g_\alpha,\frak g_\beta,\frak g_\gamma]\subseteq \frak g_{\alpha+\beta+\gamma},\;
     \alpha,\beta,\gamma\in {\mathbb Z}_2,
$$
which is graded skew-symmetric
\begin{equation}
[a,b,c] = -(-1)^{|a||b|}[b,a,c],\label{skew-symmetry 1}
\end{equation}
\begin{equation}
[a,b,c] = -(-1)^{|b||c|}[a,c,b],\label{skew-symmetry 2}
\end{equation}
\begin{equation}
[a,b,c] = -(-1)^{|a||b|+|b||c|+|a||c|}[c,b,a],\label{skew-symmetry 3}
\end{equation}
and satisfies the graded Filippov-Jacobi identity
\begin{eqnarray}
[a,b,[c,d,e]] = [[a,b,c],d,e]+(-1)^{|ab||c|}[c,[a,b,d],e]+(-1)^{|ab||cd|}[c,d,[a,b,e]].
\label{graded Filippov identity}
\end{eqnarray}
\vskip.3cm
\noindent
It is shown in \cite{Awata-Li-Minic-Yaneya} that one can define a skew-symmetric triple Lie bracket for square matrices of $n$th order by means of binary commutator of two matrices and the trace, and it is proved that this triple Lie bracket satisfies the Filippov-Jacobi identity. Thus a Lie algebra $\frak{gl}(n)$ induces the 3-Lie algebra. Now our aim is to show that the method proposed in \cite{Awata-Li-Minic-Yaneya} can be extended to 3-Lie superalgebras. By other words it will be shown that given a matrix Lie superalgebra $\frak{gl}(m,n)$ we can construct a graded skew-symmetric triple Lie bracket with the help of graded commutator of two matrices $X,Y\in \frak{gl}(m,n)$ and a supertrace. Then it will be proved that this graded triple Lie bracket satisfies the graded Filippov-Jacobi identity (\ref{graded Filippov identity}).

\vskip.3cm
\noindent
Let us consider a vector space of bloc matrices
\begin{equation}
X=\left(
    \begin{array}{cc}
      A & B \\
      C & D \\
    \end{array}
  \right),
\label{bloc matrix X}
\end{equation}
where $A$ is a square matrix of order $m$, and $D$ is a square matrix of order $n$. One can equip this vector space with a structure of super vector space by splitting it into two subspaces,
where the first subspace is the subspaces of bloc matrices (\ref{bloc matrix X}) with $B=C=0$ (matrices of even degree), and the second subspace is the subspace of matrices (\ref{bloc matrix X}) with $A=D=0$ (matrices of odd degree). In order to simplify notations we will denote the degree of a matrix by the same, but small letter. For instant, if a matrix is denoted by $X$ then its degree will be denoted by $x$. We will also denote the sum of two degrees $x,y$ by $\overline{xy}$, i.e. $\overline{xy}=x+y$. The super vector space of bloc matrices (\ref{bloc matrix X}) becomes the Lie superalgebra $\frak{gl}(m,n)$ if we equip it with the graded commutator of two matrices
\begin{equation}
[X,Y]=X\cdot Y- (-1)^{xy}\,Y\cdot X.
\label{graded commutator of matrices}
\end{equation}
Hence for any three matrices $X,Y,Z\in \frak{gl}(m,n)$ the graded Jacobi identity (\ref{graded binary Jacobi identity}) holds
\begin{equation}
[X,[Y,Z]]=[[X,Y],Z]+(-1)^{xy}[Y,[X,Z]].
\end{equation}
If at least one of matrices $X,Y$ is of even degree then the graded commutator (\ref{graded commutator of matrices}) of $X,Y$ is the usual commutator of two matrices $[X,Y]=X\cdot Y-Y\cdot X$. If $X,Y$ are matrices of odd degree then the graded commutator (\ref{graded commutator of matrices}) is anti-commutator which will be denoted by $\{X,Y\}=X\cdot Y+Y\cdot X$.
We will denote the subspace of matrices of even degree by $\frak{gl}_0(m,n)$, and the subspace of matrices of odd degree by $\frak{gl}_1(m,n)$. Evidently
$$
\frak{g}=\frak{gl}_0(m,n)\oplus\frak{gl}_1(m,n),
$$
and $\frak{gl}_0(m,n)$ is the subalgebra of $\frak{gl}(m,n)$. We remind that the supertrace of a matrix (\ref{bloc matrix X}) is defined by
\begin{equation}
\mbox{Str}\,X=\mbox{Tr}\,A-\mbox{Tr}\,D.
\end{equation}
The important property of supertrace is that supertrace of graded commutator of any two matrices is zero, i.e.
\begin{equation}
\mbox{Str}\,([X,Y])=0.
\end{equation}
It is also worth to mention that supertrace of any odd degree matrix is zero, i.e. $\mbox{Str}\,X=0$ for any $X\in \frak{gl}_1(m,n)$.

\vskip.3cm
\noindent
Now we define a graded triple Lie bracket of matrices $X,Y,Z\in\frak{gl}(m,n)$ by
\begin{eqnarray}
[X,Y,Z]=\mbox{Str}\,X\;[Y,Z]+(-1)^{\overline{yz}\,x}\mbox{Str}\,Y\;[Z,X]+ (-1)^{z\,\overline{xy}}\mbox{Str}\,Z\;[X,Y].
\label{triple Lie bracket}
\end{eqnarray}
A number of terms at the right-hand side of this triple Lie bracket depends on the parities of matrices $X,Y,Z$. Let us consider the structure of the formula of graded triple commutator for all possible combinations of parities of matrices. We begin with the matrices of even subspace. If all three matrices $X,Y,Z$ are even matrices, then the right-hand side of (\ref{triple Lie bracket}) is the sum of cyclic permutations of $X,Y,Z$ in $\mbox{Str}\,X\;[Y,Z]$, i.e.
\begin{eqnarray}
[A,B,C] =\mbox{Str}(A)\;[B,C]+\mbox{Str}(B)\;[C,A]+\mbox{Str}(C)\;[A,B].
     \label{triple graded for even matrices}
\end{eqnarray}
We see that in the even subspace we have the formula for graded triple commutator which is similar to the triple commutator (\ref{ternary commutator with trace}) used in the paper \cite{Awata-Li-Minic-Yaneya}, but it is not exactly the same if the block $D$ in a matrix (\ref{bloc matrix X}) is not trivial. Taking into account that the supertrace of a matrix of odd degree is zero, we get from (\ref{triple Lie bracket}) the expressions for graded triple commutators of matrices for other combinations of parities
\begin{enumerate}
\item if $X$ is an odd matrix and $Y,Z$ are even matrices then
\begin{equation}
[X,Y,Z] = \mbox{Str}\,Y\;[Z,X]-\mbox{Str}\,Z\;[Y,X],
\label{odd even even}
\end{equation}
\item if $X,Y$ are odd matrices and $Z$ is even then
\begin{equation}
[X,Y,Z] = \mbox{Str}\,Z\;\{X,Y\},
\label{odd odd even}
\end{equation}
\item
if $X,Y,Z$ are odd degree matrices then
$$[X,Y,Z]=0.$$
\end{enumerate}
These formulae (and similar formulae for various arrangements of odd and even matrices in triple Lie bracket) clearly show that (\ref{triple Lie bracket}) is the graded triple Lie bracket, i.e. the degree of $[X,Y,Z]$ is equal to the sum of degrees of matrices $X,Y,Z$.

\vskip.3cm
\noindent
Now our aim is to show that the graded triple Lie bracket (\ref{triple Lie bracket}) is graded skew-symmetric. Indeed rearranging matrices $X,Y$ in the graded triple Lie bracket we get
\begin{eqnarray}
[Y,X,Z]=\mbox{Str}\,Y\;[X,Z]+(-1)^{\overline{xz}\,y}\mbox{Str}\,X\;[Z,Y]+ (-1)^{z\,\overline{xy}}\mbox{Str}\,Z\;[Y,X].
\label{permutation of X,Y}
\end{eqnarray}
Making use of the graded skew-symmetry of graded commutator, each term of the right-hand side of (\ref{permutation of X,Y}) can be written as follows:
\begin{eqnarray}
(-1)^{\overline{xz}\,y}\mbox{Str}\,X\;[Z,Y] &=& -(-1)^{xy}\,\mbox{Str}\,X\;[Y,Z],\nonumber\\
\mbox{Str}\,Y\;[X,Z] &=& -(-1)^{xy}\big( (-1)^{x\,\overline{yz}}\mbox{Str}\,Y\;[Z,X]\big),\nonumber\\
(-1)^{z\,\overline{xy}}\mbox{Str}\,Z\;[Y,X] &=& -(-1)^{xy}\big((-1)^{z\,\overline{xy}}\mbox{Str}\,Z\;[X,Y]\big).\nonumber
\end{eqnarray}
Substituting the right-hand sides of these relations into (\ref{permutation of X,Y}) we obtain
$$
[X,Y,Z]=-(-1)^{xy}[Y,X,Z],
$$
and this proves the relation (\ref{skew-symmetry 1}).
Analogously we can verify the relations (\ref{skew-symmetry 2}),(\ref{skew-symmetry 3}).
\begin{theorem}
The graded triple Lie bracket (\ref{triple Lie bracket}) satisfies the graded Filippov-Jacobi identity
\begin{eqnarray}
[X,Y,[Z,V,W]] &=& [[X,Y,Z],V,W]+(-1)^{\overline{xy}\,z}[Z,[X,Y,V],W]\nonumber\\
      &&\qquad\qquad\qquad\quad +(-1)^{\overline{xy}\;\overline{zv}}[Z,V,[X,Y,W]].
      \label{graded Filippov identity theorem}
\end{eqnarray}
\end{theorem}

\noindent
In order to prove this theorem let us denote
\begin{eqnarray}
A=[Z,V,W],\;\;B=[X,Y,Z],C=[X,Y,V],\;\;D=[X,Y,W].\nonumber
\end{eqnarray}
Then the graded Filippov-Jacobi identity (\ref{graded Filippov identity theorem}) can be written as
\begin{eqnarray}
[X,Y,A]=[B,V,W]+(-1)^{\overline{xy}\,z}[Z,C,W]+(-1)^{\overline{xy}\;\overline{zv}}[Z,V,D].
\label{graded Filippov identity}
\end{eqnarray}
We begin with the left-hand side of graded Filippov-Jacobi identity. The matrix $A$ is the linear combination of graded binary commutators
\begin{eqnarray}
A=\mbox{Str}\,Z\,[V,W]+(-1)^{z\,\overline{vw}}\mbox{Str}\,V\,[W,Z]+(-1)^{w\,\overline{zv}}\mbox{Str}\,W\,[Z,V].\nonumber
\end{eqnarray}
Because the super trace makes a graded commutator vanish we have $\mbox{Str}\,A=0$. Thus the left-hand side of the Filippov-Jacobi identity (\ref{graded Filippov identity}) takes the form
\begin{eqnarray}
[X,Y,A] &=& \mbox{Str}\,X\,[Y,A]+(-1)^{x\,\overline{ya}}\mbox{Str}\,Y\,[A,X]=\nonumber\\
     &=& \mbox{Str}\,X\,\mbox{Str}\,Z\,[Y,[Z,W]]+\mbox{g.c.p.} (Z,V,W)\nonumber\\
     &&\qquad\qquad+ (-1)^{x\,\overline{ya}}\big(\mbox{Str}\,Y\,\mbox{Str}\,Z\,[[Z,W],X]\nonumber\\
     &&\qquad\qquad\qquad+\mbox{g.c.p.} (Z,V,W)\big),\label{left-hand side of identity}
\end{eqnarray}
where $a$ is the degree of a matrix $A$, and "g.c.p.$(Z,V,W)$" (graded cyclic permutations) means that one should perform a cyclic permutations of matrices $Z,V,W$ in a preceding term together with multiplication by $(-1)^{z\,\overline{vw}}$ for the permutation $(Z,V,W)\to(V,W,Z)$ and by $(-1)^{w\,\overline{zv}}$ for $(Z,V,W)\to(W,Z,V)$. Thus there are six terms at the left-hand side of the graded Filippov-Jacobi identity (\ref{graded Filippov identity}).

\vskip.3cm
\noindent
Analogously $\mbox{Str}\,B=\mbox{Str}\,C=\mbox{Str}\,D=0$, and the terms at the right-hand side of the graded Filippov-Jacobi identity (\ref{graded Filippov identity}) can be written as
\begin{eqnarray}
[B,V,W] &=& (-1)^{b\,\overline{vw}}\big(\mbox{Str}\,V\,\mbox{Str}\,X\,[W,[Y,Z]]+
\mbox{g.c.p.}(X,Y,Z)\big)\nonumber\\&&\qquad\qquad +(-1)^{w\,\overline{bv}}\big(\mbox{Str}\,W\,\mbox{Str}\,X\,[[Y,Z],V]
                                                 +\mbox{g.c.p.}(X,Y,Z)\big),\cr
[Z,C,W] &=& \mbox{Str}\,Z\,\mbox{Str}\,X\,[[Y,V],W]+
\mbox{g.c.p.}(X,Y,V)\nonumber\\ &&\qquad\qquad+(-1)^{w\,\overline{cz}}\big(\mbox{Str}\,W\,\mbox{Str}\,X\,[Z,[Y,V]]
                                                 +\mbox{g.c.p.}(X,Y,V)\big),\nonumber\cr
[Z,C,W] &=& \mbox{Str}\,Z\,\mbox{Str}\,X\,[V,[Y,W]]+
\mbox{g.c.p.}(X,Y,W)\nonumber\\ &&\qquad\qquad +(-1)^{z\,\overline{vd}}\big(\mbox{Str}\,V\,\mbox{Str}\,X\,[[Y,W],Z]
                                                 +\mbox{g.c.p.}(X,Y,W)\big).\nonumber
\end{eqnarray}
Thus there are totally 18 terms at the right-hand side of the graded Filippov-Jacobi identity.

\vskip.3cm
\noindent
Now let us consider the first term at the left-hand side of the identity (\ref{left-hand side of identity})
\begin{equation}
\mbox{Str}\,X\,\mbox{Str}\,Z\;[Y,[Z,W]].
\label{term at the left}
\end{equation}
There are two similar terms at the right-hand side of the identity, and they are the first terms in the expressions for $[B,V,W]$ and $[Z,C,W]$ multiplied by corresponding coefficients shown in (\ref{graded Filippov identity}))
\begin{equation}
\mbox{Str}\,Z\,\mbox{Str}\,X\,\big((-1)^{z\,\overline{xy}}[[Y,V],W]+
     (-1)^{\overline{zv}\;\overline{xy}}[V,[Y,W]]\big).
     \label{two terms at the right}
\end{equation}
The terms (\ref{term at the left}),(\ref{two terms at the right}) vanish if at least one of matrices $X,Z$ is of odd degree, because the super trace of a matrix of odd degree is zero. Hence what is remained is the case when both matrices $X,Z$ are matrices of even degree that is $x=z=0$. But in this case the expression (\ref{two terms at the right}) takes the form
\begin{equation}
\mbox{Str}\,Z\,\mbox{Str}\,X\,\big([[Y,V],W]+
     (-1)^{vy}[V,[Y,W]]\big).
\label{two terms for graded Jacobi}
\end{equation}
We see that the term (\ref{term at the left}) at the left-hand side of the identity is canceled by two terms (\ref{two terms for graded Jacobi}) of the right-hand side of identity because all together they form the usual graded Jacobi identity. Analogously it can be shown that all the rest five terms at the left-hand side of the identity (\ref{left-hand side of identity}) are canceled by the corresponding terms of the right-hand side of the identity.

\vskip.3cm
\noindent
Now after these cancellations there are no more non-trivial terms at the left-hand side of the identity, and at the right-hand side there are $18-12=6$ terms. These terms can be split into pairs such that they cancel each other. For instant, in the first half of the expression $[B,V,W]$ there is the term
\begin{equation}
(-1)^{b\,\overline{vw}+z\,\overline{xy}}\big(\mbox{Str}\,V\,\mbox{Str}\,Z\,[W,[X,Y]],
\end{equation}
where $b=x+y+z$, determined by the permutation $(X,Y,Z)\to (Z,X,Y)$. This term is non-trivial only when $v=z=0$, i. e.
\begin{equation}
(-1)^{w\,\overline{xy}}\big(\mbox{Str}\,V\,\mbox{Str}\,Z\,[W,[X,Y]].
\label{one term in the right}
\end{equation}
The expression $[Z,C,W]$, accordingly to (\ref{graded Filippov identity}) multiplied by $(-1)^{z\,\overline{xy}}$, has the similar term
$$
(-1)^{\overline{vz}\;\overline{xy}}\mbox{Str}\,V\,\mbox{Str}\,Z\,[[X,Y],W],
$$
which is determined by the cyclic permutation $(X,Y,V)\to (V,X,Y)$ in the first half of the expression $[Z,C,W]$. Because this term is non-trivial only in the case $v=z=0$, we can write it as follows
\begin{equation}
\mbox{Str}\,V\,\mbox{Str}\,Z\,[[X,Y],W]=
         -(-1)^{w\;\overline{xy}}\mbox{Str}\,V\,\mbox{Str}\,Z\,[W,[X,Y]].
\label{second term in the right}
\end{equation}
Now it is evident that the sum of two terms (\ref{one term in the right}),(\ref{second term in the right}) is zero, and this ends the proof.

\vskip.3cm
\noindent
This theorem shows that a vector space of bloc matrices (\ref{bloc matrix X}) equipped with the graded triple Lie bracket (\ref{triple Lie bracket}) is the 3-Lie superalgebra, and we will denote this 3-Lie superalgebra by $\frak{glt}(m,n)$. Evidently this algebra can be viewed as the 3-Lie superalgebra induced by the matrix Lie superalgebra $\frak{gl}(m,n)$, because we construct the graded triple Lie bracket (\ref{triple Lie bracket}) by means of graded binary commutator of two bloc matrices.
\section{3-Lie superalgebras $\frak{glt}(1,1), \frak{glt}(2,2)$ . Pauli and Dirac matrices}
In this section we study the structure of induced matrix 3-Lie superalgebra with the help of generators and the structure constants. We study the structure of the 3-Lie superalgebra $\frak{glt}(1,1)$ of $2\times 2$-matrices choosing the Pauli matrices and the unit matrix $E_2$ as the basis for the vector space of $2\times 2$-matrices, and the structure of the 3-Lie superalgebra $\frak{glt}(2,2)$ of $4\times 4$-matrices taking the Dirac matrices and the unit matrix $E_4$ as the basis for the vector space of $4\times 4$-matrices. We find the non-trivial graded triple commutators of these algebras.

\vskip.3cm
\noindent
Let us consider the matrix 3-Lie superalgebra $\frak{glt}(m,n)$ induced by a matrix Lie superalgebra $\frak{gl}(m,n)$.
Let us choose a basis $T_i$, where $i=1,2,\ldots,m^2+n^2$, for the even subspace $\frak{gl}_0(m,n)$ and a basis $S_\mu$, where $\mu=1,2,\ldots,2mn$, for the odd subspace $\frak{gl}_1(m,n)$. Let us denote the structure constants of a matrix Lie superalgebra $\frak{gl}(m,n)$ as follows:
\begin{equation}
[T_i,T_j]=c^k_{ij}\;T_k,\;\;
               [T_i,S_\mu]=\zeta^\nu_{i\mu}\;S_\nu,\;\;
                     \{S_\mu,S_\nu\}=\eta^i_{\mu\nu}\;T_i.
\end{equation}
The structure constants of the induced matrix 3-Lie superalgebra will be denoted by
\begin{eqnarray}
[T_i,T_j,T_k] &=&\tilde c^{\,l}_{ijk}\;T_l,\;\;\;\;\;\;\;
            [S_\mu,T_i,T_j]=\tilde\zeta^{\,\nu}_{\mu ij}\;S_\nu,\nonumber\\
            &&\;\;\;[S_\mu,S_\nu, T_i] =\tilde\eta^{\,j}_{\mu\nu i}\;T_j.\nonumber
\end{eqnarray}
The structure constants in the case of a graded triple commutator of odd generators $S_\mu$ are all zeros, because the supertrace of any odd degree matrix is zero, i.e. $[S_\mu,S_\nu,S_\sigma]=0$.

\vskip.3cm
\noindent
Let us denote the supertrace of even degree generators of $\frak{gl}(m,n)$ by $s_i=\mbox{Str}\,T_i$. Then we can find that the structure constants of the induced matrix 3-Lie superalgebra can be expressed in terms of the structure constants of a matrix Lie superalgebra $\frak{gl}(m,n)$ as follows:
\begin{eqnarray}
\tilde c^{\,l}_{ijk} &=& s_i\,c^l_{jk}+s_j\,c^l_{ki}+s_k\,c^l_{ij},\nonumber\\
\tilde\zeta^{\,\nu}_{\mu ij} &=& s_i\,\zeta^{\nu}_{j\mu}-s_j\,\zeta^{\nu}_{i\mu},\label{structure constants}\\
\tilde\eta^{\,j}_{\mu\nu i} &=& s_i\,\eta^j_{\mu\nu}.\nonumber
\end{eqnarray}
Let us consider the Lie superalgebra $\frak{gl}(1,1)$. We can choose the basis for the vector space of $\frak{gl}(1,1)$ which consists of the unit matrix of second order $E_2$ and Pauli matrices $\sigma_1,\sigma_2,\sigma_3$, where
$$
\sigma_1=\left(
           \begin{array}{cc}
             0 & 1 \\
             1 & 0 \\
           \end{array}
         \right),\;\; \sigma_2=\left(
                                 \begin{array}{cc}
                                   0 & -i \\
                                   i & 0 \\
                                 \end{array}
                               \right),\;\; \sigma_3=\left(
                                              \begin{array}{cc}
                                                1 & 0 \\
                                                0 & -1 \\
                                              \end{array}
                                            \right).
$$
The matrix $\sigma_3=i^{-1}\sigma_1\,\sigma_2$ determines the super vector space structure of $ \mathbb C^2$, where the eigenvectors with eigenvalue 1 (-1) form the subspace of even (odd) vectors $\mathbb C^2_0$ ($\mathbb C^2_1$), and $\mathbb C=\mathbb C^2_0\oplus\mathbb C^2_1$. Then, according to this super vector space structure of $\mathbb C^2$, the matrices $E_2,\sigma_3$ are the even generators of the Lie superalgebra $\frak{gl}(1,1)$ and the Pauli matrices $\sigma_1,\sigma_2$ are the odd generators of $\frak{gl}(1,1)$. We have the following non-trivial graded commutators of this Lie superalgebra:
\begin{eqnarray}
[\sigma_3,\sigma_1] &=& 2i\,\sigma_2,\;\;[\sigma_3,\sigma_2]=-2i\,\sigma_1,\\
\{\sigma_1,\sigma_1\} &=& 2\,E_2,\;\;\; \{\sigma_2,\sigma_2\}=2\, E_2.\nonumber
\end{eqnarray}
If we denote the generators of Lie superalgebra $\frak{gl}(1,1)$ by $e_1=E_2,e_2=\sigma_3$ and $f_{\bar 1}=\sigma_1,f_{\bar 2}=\sigma_2,$, then the non-trivial structure constants are
\begin{equation}
\zeta^{\bar 2}_{2\bar 1}=2i,\;\;\zeta^{\bar 1}_{2\bar 2}=-2i,\;\;\;\;\eta^{1}_{\bar 1\bar 1}=\eta^{1}_{\bar 2\bar 2}=2.
\end{equation}
We have $s_1=\mbox{Str}\,E_2=0,s_2=\mbox{Str}\,\sigma_3=2$ and, making use of (\ref{structure constants}), we find
the non-trivial structure constants of 3-Lie superalgebra $\frak{glt}(1,1)$
\begin{equation}
\tilde\eta^1_{\bar 1\bar 1 2}=s_2\,\eta^1_{\bar 1\bar 1}=4,\;\;
   \tilde\eta^1_{\bar 2\bar 2 2}=s_2\,\eta^1_{\bar 2\bar 2}=4.
\end{equation}
Thus the 3-Lie superalgebra $\frak{glt}(1,1)$ is generated by two even generators $e_1,e_2$, two odd generators $f_{\bar 1},f_{\bar 2}$ and the non-trivial triple commutators of this algebra are
\begin{equation}
[f_{\bar 1},f_{\bar 1},e_2]=4\,e_1,\;\;\;[f_{\bar 2},f_{\bar 2},e_2]=4\,e_1.
\end{equation}
By other words, the Pauli matrices $\sigma_1,\sigma_2,\sigma_3$ and the unit matrix $E_2$ generate the 3-Lie superalgebra $\frak{glt}(1,1)$, where $E_2,\sigma_3$ are the even generators, $\sigma_1,\sigma_2$ are the odd generators, and the non-trivial graded triple commutators are
\begin{equation}
[\sigma_1,\sigma_1,\sigma_3]=4\,E_2,\;\;[\sigma_2,\sigma_2,\sigma_3]=4\,E_2.
\end{equation}

\vskip.3cm
\noindent
Let us consider the 3-Lie superalgebra $\frak{glt}(2,2)$ of $4\times 4$-matrices induced by the Lie superalgebra $\frak{gl}(2,2)$. We choose the Dirac matrices
\begin{eqnarray}
\gamma_0 = \left(
           \begin{array}{cc}
             0 & -i\,E_2 \\
             i\,E_2 & 0 \\
           \end{array}
         \right),\;\;
\gamma_a=\left(
           \begin{array}{cc}
             0 & \sigma_a \\
             \sigma_a & 0 \\
           \end{array}
         \right),\label{Dirac matrices}
\end{eqnarray}
where $a=1,2,3$, and the unit $4\times 4$-matrix $E_4$ as the basis for the vector space of $\frak{gl}(2,2)$. The Dirac matrices satisfy the relations
\begin{equation}
\gamma_\mu\gamma_\nu+\gamma_\nu\gamma_\mu=2\,\delta_{\mu\nu}\,E_4.
\end{equation}
The matrix
$$
\gamma_5=\gamma_0\gamma_1\gamma_2\gamma_3=\left(
                                            \begin{array}{cc}
                                              E_2 & 0 \\
                                              0 & -E_2 \\
                                            \end{array}
                                          \right),
$$
defines the super vector space structure of $\mathbb C^4$, and the Dirac matrices (\ref{Dirac matrices}) are the odd matrices with respect to this super vector space structure of $\mathbb C^4$. Let us denote
$$
\gamma_{\mu_1\ldots\mu_k}=\gamma_{\mu_1}\ldots\gamma_{\mu_k}.
$$
It is also useful to introduce the four matrices $\gamma^\ast_\mu$ by
\begin{equation}
\gamma^\ast_\mu=\frac{1}{3!}\epsilon_{\mu\nu\sigma\tau}\;\gamma_{\nu\sigma\tau},
\end{equation}
where $\epsilon_{\mu\nu\sigma\tau}$ is the totally anti-symmetric tensor in four dimensional vector space. Then
\begin{equation}
\gamma^\ast_0=\left(
           \begin{array}{cc}
             0 & i\,E_2 \\
             i\,E_2 & 0 \\
           \end{array}
         \right),\;\;\;
\gamma^\ast_a=\left(
           \begin{array}{cc}
             0 & -\sigma_a \\
             \sigma_a & 0 \\
           \end{array}
         \right).
\end{equation}
Hence the matrix 3-Lie superalgebra $\frak{glt}(2,2)$ is generated by the following 16 matrices:
\begin{eqnarray}
\mbox{even subspace}\;\;
&\frak{glt}_0(2,2)&:\;\;E_4,\;\gamma_{\mu\nu},\;\gamma_5\;\;(\mu<\nu),\label{generators 2,2 I}\\
\mbox{odd subspace}\;\;&\frak{glt}_1(2,2)&:\;\;\gamma_{\mu},\;\,\gamma^\ast_{\nu}.
\label{generators of 2,2 II}
\end{eqnarray}
A super trace of a generator of $\frak{glt}(2,2)$ can be non-trivial only in the case of even generator. It is easy to find that
\begin{equation}
\gamma_0\gamma_a=-i\;\left(
                   \begin{array}{cc}
                     \sigma_a & 0 \\
                     0 & -\sigma_a \\
                   \end{array}
                 \right),\;\;\;
\gamma_a\gamma_b=i\,\epsilon_{abc}\,\left(
                                      \begin{array}{cc}
                                        \sigma_c & 0 \\
                                        0 & \sigma_c \\
                                      \end{array}
                                    \right),
\end{equation}
where $\epsilon_{abc}$ is the totally anti-symmetric tensor in a 3-dimensional space. The Pauli matrices are traceless matrices which implies that the super trace of any even generator $\gamma_{\mu\nu}$ vanishes. Because $\mbox{Str}\,E_4=0$, we see that the only even generator with non-trivial super trace is $\gamma_5$, and $\mbox{Str}\,\gamma_5=4$.

\vskip.3cm
\noindent
Now our aim is to find all non-trivial graded triple commutators of generators (\ref{generators 2,2 I}), (\ref{generators of 2,2 II}) of $\frak{glt}(2,2)$ composed with the help of ordered products of the Dirac matrices. Let us begin with the case of triple graded commutator of type $[\mbox{odd},\mbox{odd},\mbox{even}]$. From previous considerations it is clear that the non-trivial graded triple commutators of this type have the form $[\gamma_\mu,\gamma_\nu,\gamma_5]$, $[\gamma_\mu,\gamma^\ast_\nu,\gamma_5]$ and $[\gamma^\ast_\mu,\gamma^\ast_\nu,\gamma_5]$. We find
\begin{eqnarray}
[\gamma_\mu,\gamma_\nu,\gamma_5] &=& 8\,\delta_{\mu\nu}\,E_4,
   \;\;\;\;[\gamma^\ast_\mu,\gamma^\ast_{\nu},\gamma_5]=-8\,\delta_{\mu\nu}\,E_4,\\
&& [\gamma_\mu,\gamma^\ast_\nu,\gamma_5] =
      -8\,{\epsilon_{\mu\nu\sigma\tau}}\,\gamma_{\sigma\tau},
      \label{odd*odd*gamma}
\end{eqnarray}
where $\mu<\nu<\sigma<\tau$ in (\ref{odd*odd*gamma}).
Next we consider the case of triple graded commutators of type $[\mbox{odd},\mbox{even},\mbox{even}]$. The non-trivial graded triple commutators have the form $[\gamma_\mu,\gamma_{\sigma\tau},\gamma_5]$, $[\gamma^\ast_\mu,\gamma_{\sigma\tau},\gamma_5]$, and we find
\begin{eqnarray}
[\gamma_\mu,\gamma_{\sigma\tau},\gamma_5] &=& 8\,(\delta_{\mu\sigma}\,\gamma_\tau-\delta_{\mu\tau}\,\gamma_\sigma),\cr
[\gamma^\ast_\mu,\gamma_{\sigma\tau},\gamma_5] &=& 8\,(\delta_{\mu\sigma}\,\gamma^\ast_\tau-\delta_{\mu\tau}\,\gamma^\ast_\sigma).
\end{eqnarray}
Finally the subspace of even elements $\frak{glt}_0(2,2)$ is the subalgebra of 3-Lie superalgebra $\frak{glt}(2,2)$, and the non-trivial triple commutators of this subalgebra is given by
\begin{equation}
[\gamma_{\mu\nu},\gamma_{\sigma\tau},\gamma_5]=
  8\,(\delta_{\nu\sigma}\,\gamma_{\mu\tau}+\delta_{\nu\tau}\,\gamma_{\sigma\mu}-
    \delta_{\mu\sigma}\,\gamma_{\nu\tau}-\delta_{\mu\tau}\,\gamma_{\sigma\nu}).
\end{equation}
\section{Clifford algebra approach to 3-Lie superalgebra $\frak{glt}(n,n)$}
In this section we generalize the matrix 3-Lie superalgebras $\frak{glt}(1,1),\frak{glt}(2,2)$ by means of Clifford algebra with even number of generators.

\vskip.3cm
\noindent
Let $C_n$ be a Clifford algebra over $\mathbb C$ generated by $\gamma_1,\gamma_2,\ldots,\gamma_n$, where $n=2m$. Thus the generators of $C_n$ satisfy the Clifford algebra commutation relations
$$
\gamma_i\gamma_j+\gamma_j\gamma_i=2\,\delta_{ij}\,\mathbb 1,
$$
where $\mathbb 1$ is e unit element of $C_n$. Let $N=\{1,2,\ldots,n\}$, $I=\{i_1,i_2,\ldots,i_k\}$, where $1\leq i_1<i_2<\ldots<i_k\leq n$, be a subset of $N$ and $|I|$ be the number of integers in $I$, i.e. $|I|=k$. For any two subsets $I,J\subset N$ we define the integer
$$
\epsilon(I,J)=\sum_{j\in J}\epsilon(I,j),
$$
where $\epsilon(I,j)$ is the number of integers in $I$, which are greater than $j$.
Let us associate to each subset $I$ of $N$ a Clifford algebra monomial $\gamma_I$ as follows $\gamma_I=\gamma_{i_1}\gamma_{i_2}\ldots\gamma_{i_k}, \gamma_{\emptyset}=\mathbb 1$. Then the vector space of Clifford algebra $C_n$ is spanned by these monomials and any element of Clifford algebra can be expressed as a linear combination of monomials of basis
$$
\gamma=\sum_{I\subset N} \xi_I\,\gamma_I,
$$
where $\xi_I=\xi_{i_1i_2\ldots i_k}$ is a complex number.

\vskip.3cm
\noindent
Clifford algebra $C_n$ is the superalgebra if one associates to each monomial $\gamma_I$ the degree $|\gamma_I|$ as follows $|\gamma_I|=|I|\; (\mbox{mod}\,2)$, i.e. $\gamma_I$ is even degree monomial if it is a product of even number of generators, and $\gamma_I$ is odd degree monomial if it contains the odd number of generators. If we denote by $C_n^0$ the subspace of even degree monomials and by $C_n^1$ the subspace of odd degree monomials then
$$
C_n=C_n^0\oplus C_n^1.
$$
A product of two monomials of basis of Clifford algebra $\gamma_I,\gamma_J$ can be written as monomial of basis $\gamma_I\gamma_J=(-1)^{\epsilon(I,J)}\gamma_{I\Delta J}$, where
$I\Delta J=(I\cup J)\setminus (I\cap J)$ is the symmetric difference of two subsets. For symmetric difference of two subsets we have the obvious property $|I|+|J|=|I\Delta J|\;(\mbox{mod}\,2)$. Hence for any two monomials of basis it holds $|\gamma_I\gamma_J|=|\gamma_I|+|\gamma_J|$ which means that Clifford algebra $C_n$ is the superalgebra.

\vskip.3cm
\noindent
The Lie superalgebra structure of the superalgebra $C_n$ is determined by the graded commutator $[\gamma,\gamma^\prime]=\gamma\gamma^\prime-(-1)^{|\gamma||\gamma^\prime|}\gamma^\prime\gamma$, where $\gamma,\gamma^\prime\in C_n$ are homogeneous elements of superalgebra $C_n$. Then the monomials of basis $\gamma_I$ can be considered as the generators of this Lie superalgebra, and for any two generators $\gamma_I,\gamma_J$ we have
\begin{equation}
[\gamma_I,\gamma_J]=f(I,J)\;\gamma_{I\Delta J},
\label{binary commutators}
\end{equation}
where $f(I,J)$ is the integer-valued function of two subsets of $N$ defined by
$$
f(I,J)=(-1)^{\epsilon(I,J)}\big(1-(-1)^{|I\cap J|}\big).
$$
The Lie superalgebra of a Clifford algebra $C_n$ is isomorphic to the matrix Lie superalgebra $\frak{gl}(n,n)$. Let $\mathbb C^2$ be the super vector space with $\mathbb Z_2$-graded structure defined by the Pauli matrix $\sigma_3$. If we identify the generators $\gamma_1,\gamma_2$ of the Clifford algebra $C_2$ with the Pauli matrices $\gamma_1=\sigma_1,\gamma_2=\sigma_2$ then the Lie superalgebra of $C_2$ is isomorphic to the matrix Lie superalgebra $\frak{gl}(1,1)$. Taking the graded tensor product $\mathbb C^n=\mathbb C^2\otimes\mathbb C^2\otimes\ldots\otimes\mathbb C^2 (m\;\mbox{times})$ and identifying generators $\gamma_{2j-1},\gamma_{2j}$, where $j=1,2,\ldots,m$, of Clifford algebra $C_n$ with $\sigma_1,\sigma_2$ in $j$th factor of this graded tensor product, we get $C_n\simeq \mbox{End}\,\mathbb C^n$. From this it follows that the Lie superalgebra of $C_n$ is isomorphic to the matrix Lie superalgebra $\frak{gl}\,(n,n)$.

\vskip.3cm
\noindent
Now making use of the isomorphism between the Lie superalgebra of $C_n$ and $\frak{gl}\,(n,n)$ we can consider the super trace of any element of a Clifford algebra $C_n$. Hence we can construct the 3-Lie superalgebra of a Clifford algebra $C_n$ by means of the graded triple commutator
\begin{eqnarray}
[\gamma_1,\gamma_2,\gamma_3] &=& \mbox{Str}\,\gamma_1\;[\gamma_2,\gamma_3]
        +(-1)^{|\gamma_1|(|\gamma_2|+|\gamma_3|)}\mbox{Str}\,\gamma_2\;[\gamma_3,\gamma_1]\nonumber\\                                                                   &&\qquad\qquad\qquad +(-1)^{|\gamma_3|(|\gamma_1|+|\gamma_2|)}\mbox{Str}\,\gamma_3\;[\gamma_1,\gamma_2],
\end{eqnarray}
where $\gamma_1,\gamma_2,\gamma_3\in C_n$. We can find the non-trivial commutators of the monomials of basis of $C_n$ which generalize the 3-Lie superalgebras induced by Pauli and Dirac matrices. First it follows from the way the isomorphism between the Lie superalgebra of $C_n$ and $\frak{gl}\,(n,n)$ is constructed that the super trace of any monomial $\gamma_I$, where $I\subset I, I\neq N$ is zero. This can be easily proved by means of the formula
\begin{equation}
\mbox{Str}\,(L\otimes K)=\mbox{Str}\,L\;\mbox{Str}\,K,
\label{super trace of tensor product}
\end{equation}
where $L:V\to V, K:W\to W$ are linear operators of super vector spaces $V,W$ and $L\otimes K$ is their graded tensor product. If $\gamma_I$ is a monomial of basis such that $I\neq N$ then in the corresponding graded tensor product of linear operators of the super vector space $\mathbb C^2$ there is at least one $j$th factor which contains either $\sigma_1$ or $\sigma_2$ (not their product). But $\mbox{Str}\,\sigma_1=\mbox{Str}\,\sigma_2=0$, and (\ref{super trace of tensor product}) implies that the super trace of the whole product vanishes. Second $\mbox{Str}\,\sigma_1\sigma_2=i\;\mbox{Str}\,\sigma_3=2i$, and the formula (\ref{super trace of tensor product}) immediately gives
\begin{equation}
\mbox{Str}\,\gamma_N=(2i)^m.
\end{equation}
Thus the non-trivial graded commutators of the 3-Lie superalgebra of $C_n$ have the form $[\gamma_I,\gamma_J,\gamma_N]$, and applying the formulae (\ref{odd even even}), (\ref{odd odd even}), (\ref{binary commutators}) we find
\begin{equation}
[\gamma_I,\gamma_J,\gamma_N]=\mbox{Str}\,\gamma_N\;[\gamma_I,\gamma_J]=(2i)^m\,f(I,J)\,\gamma_{I\Delta J}.
\end{equation}
The 3-Lie superalgebras of the Pauli and Dirac matrices are the particular cases of 3-Lie superalgebra of $C_n$.
\section{3-Lie superalgebra approach to BRST-operator}
We wish to conclude the present paper by pointing out a possible application of a 3-Lie superalgebra $\frak{glt}(m,n)$ of matrices in a gauge field theory. We remind that in the paper \cite{Awata-Li-Minic-Yaneya} the authors define the triple commutator of square matrices of $N$th order
$$
[A,B,C]=\mbox{Tr}\,A\;[B,C]+\mbox{Tr}\,B\;[C,A]+\mbox{Tr}\,C\;[A,B],
$$
which induces the 3-Lie algebra of $N\times N$-matrices, and they consider an analog of gauge transformation of a matrix $A$
\begin{equation}
\delta A=i[X,Y,A].
\label{gauge transformation delta A 1}
\end{equation}
It is easy to write this gauge transformation in the form
\begin{equation}
\delta A=i\,\big([(\mbox{Tr}\,X)\,Y-(\mbox{Tr}\,Y)\,X,A]+\mbox{Tr}\,A\;[X,Y]\big).
\label{gauge transformation delta A 2}
\end{equation}
Evidently $\mbox{Tr}\big(\,(\mbox{Tr}\,X)\,Y-(\mbox{Tr}\,Y)\,X\big)=0$. Hence the first term in (\ref{gauge transformation delta A 2}) can be viewed as the usual $\mbox{su}\,(N)$ infinitesimal gauge transformation, while the second term is something new.

\vskip.3cm
\noindent
As we propose a 3-Lie superalgebra $\frak{glt}(m,n)$ of matrices constructed with the help of the graded triple commutator (\ref{triple Lie bracket}), we can consider an analog of supersymmetry transformations in gauge theory. Let $X, A,\in \frak{glt}_0(m,n)$ be two even matrices and $\Lambda,\Psi\in \frak{glt}_1(m,n)$ be two odd matrices. We can consider the following analog of supersymmetry transformation
\begin{equation}
\delta A=[X,\Lambda,A],\quad \delta\Psi=[X,\Lambda,\Psi],
\label{definition of supersymmetry transformation}
\end{equation}
where we use the graded triple commutator (\ref{triple Lie bracket}) and consider $X,\Lambda$ as parameters of supersymmetry transformation of "bosonic field" $A$ and a "fermionic field" $\Psi$. Making use of (\ref{triple Lie bracket}), we can write the supersymmetry transformation of $A$ as follows
\begin{eqnarray}
\delta A &=& \mbox{Str}\,X\;[\Lambda,A]+\mbox{Str}\,\Lambda\;[A,X]+\mbox{Str}\,A\;[X,\Lambda]\nonumber\\
      &&\qquad\qquad=\mbox{Str}\,X\;[\Lambda,A]+\mbox{Str}\,A\;[X,\Lambda],
      \label{A transformation}
\end{eqnarray}
where the middle term $\mbox{Str}\,\Lambda\;[A,X]$ vanishes because $\Lambda$ is a fermionic field and $\mbox{Str}\,\Lambda=0$. Now the first term $\mbox{Str}\,X\;[\Lambda,A]$ is similar to usual BRST-supersymmetry transformation $\delta A_\mu=\partial_\mu c+[A_\mu,c]$ in a gauge field theory \cite{Faddeev-Slavnov}, where $c$ is a ghost field, which takes the form $\delta A_\mu=-[c,A_\mu]$ if a ghost field $c$ does not depend on a point of a manifold. Apparently the second term $\mbox{Str}\,A\;[X,Y]$ is something new due to a ternary structure of a commutator. The supersymmetry transformation of a fermionic field $\Psi$ can be written
\begin{equation}
\delta\Psi=\mbox{Str}\,X\;\{\Lambda,\Psi\}+\mbox{Str}\,\Lambda\;[\Psi,X]-\mbox{Str}\,\Psi\;[X,\Lambda],
\end{equation}
and there is only one non-trivial term $\mbox{Str}\,X\;\{\Lambda,\Psi\}$ in the expression at the right-hand side because $\Lambda,\Psi$ are fermionic fields and $\mbox{Str}\,\Psi=\mbox{Str}\,\Lambda=0$. Hence
\begin{equation}
\delta\Psi=\mbox{Str}\,X\;\{\Lambda,\Psi\}.
\label{Psi transformation}
\end{equation}
If we wish to get a closed algebra of BRST-like transformation then we can identify $\Lambda\equiv \Psi$, and the transformations (\ref{A transformation}), (\ref{Psi transformation}) take on the form
\begin{eqnarray}
\delta A &=& \mbox{Str}\,X\;[\Psi,A]-\mbox{Str}\,A\;[\Psi,X],\label{BRST transformation for A}\\
\delta\Psi &=& \mbox{Str}\,X\;\{\Psi,\Psi\}.\label{BRST transformation for Psi}
\end{eqnarray}
Now the supersymmetry transformation of a fermionic field $\Psi$ is similar to BRST-transformation of a ghost field $c$ which is $\delta c=-\frac{1}{2}[c,c]$.

\vskip.3cm
\noindent
Due to a ternary structure of a commutator our supersymmetry transformation $\delta$ has two parameters $X,\Lambda$, where $X$ is a bosonic field and $\Lambda$ is a fermionic field. In order to get a closed BRST-like algebra of supersymmetry transformation we identified $\Lambda\equiv \Psi$. For $X$ we find
$$
\delta X=[X,\Psi,X]=0,
$$
and hence it is natural to identify $X$ with an auxiliary bosonic field $B$ of the usual BRST-transformation, i.e. $X\equiv B$. It is worth to mention that there is no need to introduce an auxiliary field $B$ "by hand" in order to close the BRST-algebra because this field is a part of supersymmetry transformation from the very beginning and $\delta B=0$ follows from the structure of supersymmetry transformation. Thus the supersymmetry transformation takes the form
\begin{eqnarray}
\delta A &=& \mbox{Str}\,B\;[\Psi,A]-\mbox{Str}\,A\;[\Psi,B],\nonumber\\
\delta\Psi &=& \mbox{Str}\,B\;\{\Psi,\Psi\}.\nonumber\\
\delta B &=& 0.\nonumber
\end{eqnarray}
In order to find the square of the operator $\delta$ (\ref{definition of supersymmetry transformation}) we apply the graded Filippov identity (\ref{graded Filippov identity theorem}). Then
\begin{eqnarray}
\delta^2 A&=&[X,\Lambda,[X,\Lambda,A]]=[[X,\Lambda,X],\Lambda,A]\nonumber\\
   &&\qquad\;+[X,[X,\Lambda,\Lambda],A]-[X,\Lambda,[X,\Lambda,A]].
\end{eqnarray}
The first term at the right-hand side of the above equation vanishes because $X$ is an even matrix and $[X,\Lambda,X]=0$. The last term at the right-hand side can be written as $-[X,\Lambda,[X,\Lambda,A]]=-\delta^2A$. Similar calculations in the case of a "fermionic field" $\Psi$ give
\begin{eqnarray}
\delta^2 A &=& \frac{1}{2}[X,[X,\Lambda,\Lambda],A]=\frac{1}{2}[X,\delta \Lambda,A],\\
\delta^2 \Psi &=& \frac{1}{2}[X,[X,\Lambda,\Lambda],\Psi]=\frac{1}{2}[X,\delta \Lambda,\Psi].
\end{eqnarray}
Making use of the definition of graded triple commutator (\ref{triple Lie bracket}) and identifying $X\equiv B,\Lambda\equiv \Psi$, we can write the second equation as
$$
\delta^2 \Psi=\frac{1}{2}\big( \mbox{Str}\,B\;[\delta\Psi,\Psi]+\mbox{Str}\,\delta\Psi\;[\Psi,B]+\mbox{Str}\,\Psi\;[B,\delta\Psi]\big).
$$
The last term of this expression vanishes because $\Psi$ is a fermionic field and $\mbox{Str}\,\Psi=0$. The second term vanishes because the super trace "kills" graded commutator $\mbox{Str}\,\delta\Psi=\mbox{Str}\,X\;\mbox{Str}\,\{\Psi,\Psi\}=0$. Thus
$$
\delta^2\Psi=\frac{1}{2}\mbox{Str}\,B\;[\{\Psi,\Psi\},\Psi].
$$
But it follows from the usual (binary) graded Jacobi identity that $[\{\Psi,\Psi\},\Psi]=0$, and we finally get $\delta^2\Psi\equiv 0.$

\vskip.3cm
\noindent
In the case of a bosonic field $A$ the square of $\delta$ can be written as
$$
\delta^2 A=\frac{1}{2}\big(\mbox{Str}\,B\;[\{\Psi,\Psi\},A]-\mbox{Str}\,A\;[\{\Psi,\Psi\},B]\big).
$$
Making use of the graded binary Jacobi identity, we can write
\begin{eqnarray}
[\{\Psi,\Psi\},A] &=& \{\Psi,[\Psi,A]\}+\{[\Psi,A],\Psi\}=2\,\{\Psi,[\Psi,A]\},\nonumber\cr
[\{\Psi,\Psi\},B] &=& 2\,\{\Psi,[\Psi,B]\}.\nonumber
\end{eqnarray}
Thus
\begin{eqnarray}
\delta^2 A = \{\Psi,\mbox{Str}\,B\;[\Psi,A]-\mbox{Str}\,A\;[\Psi,B]\}= \{\Psi,\delta A\}.\nonumber
\end{eqnarray}
Consequently if we assume that $\{\Psi,\delta A\}=0$ then $\delta^2\equiv 0$. It should be pointed out that this assumption is very natural because in the case of the usual BRST-operator of gauge theory $\delta A_\mu$ and a ghost field $c$ anticommute and $\{A_\mu,c\}=0$. In our case we simply assume that $\delta A$ and $\Psi$ are two odd degree matrices which anticommute.
\section*{Acknowledgments}
The author gratefully appreciates the financial support of Estonian Ministry of Education and Research by the institutional research funding IUT20-57.

\end{document}